\documentclass[twocolumn,showpacs,preprintnumbers,amsmath,amssymb]{revtex4}
\usepackage{graphicx}% Include figure files
\usepackage{dcolumn}% Align table columns on decimal point
\usepackage{bm}% bold math

\begin{document}

\preprint{APS/123-QED}

\title{
Extremely high sensitivity to uniaxial stress\\
in pressure induced superconductivity of BaFe$_2$As$_2$
}

\author{
Takehiro Yamazaki$^{1}$, Nao Takeshita$^{2,3}$, Ryosuke Kobayashi$^{1}$, 
Hideto Fukazawa$^{1,3}$, Yoh Kohori$^{1,3}$, Kunihiro Kihou$^{3}$, Chul-Ho Lee$^{1,3}$,
Hijiri Kito$^{1,3}$, Akira Iyo$^{1,3}$ and Hiroshi Eisaki$^{1,3}$
}
\address{
$^{1}$ Department of Physics, Graduate School of Science, Chiba University, Chiba 263-8522, Japan\\
$^{2}$ National Institute of Advanced Industrial Science and Technology, Tsukuba 305-8562, Japan\\
$^{3}$ JST, TRIP, Chiyoda-ku, Tokyo 102-0075, Japan\\
}
\date{\today}
\begin{abstract}
We have performed electrical resistivity measurements on single crystal BaFe$_2$As$_2$ under high pressure $P$ 
up to 16~GPa with a cubic anvil apparatus, and up to 3~GPa with a modified Bridgman anvil cell. 
The samples were obtained from the same batch, which was grown with a self-flux method. 
A cubic anvil apparatus provides highly hydrostatic pressure, and a modified Bridgman anvil cell, 
which contains liquid pressure transmitting medium, provides quasi hydrostatic pressure. 
For highly hydrostatic pressure, the crystal phase and magnetic transition temperature decreases robustly with $P$ and disappears at around 10~GPa.
The superconducting phase appears adjacent to magnetic phase in narrow pressure region between 11 and 14~GPa. 
The tiny difference of hydrostaticity between the cubic anvil apparatus and modified Bridgman anvil cell 
induces a drastic effect on the phase diagram of BaFe$_2$As$_2$. 
This result indicates that small uniaxial stress along $c$-axis strongly suppresses 
the structural/antiferromagnetic ordering and stabilizes superconductivity at much lower pressure. 
\end{abstract}
\maketitle
%
%\section{Introduction}

The discovery of superconductivity in F-doped LaFeAsO at 26~K by Kamihara $et~ al.$ heralded 
a new era of superconductivity based on iron \cite{kamihara}.
The common feature of the new superconductors is existence of iron pnictide layers, 
which is analogous to CuO$_2$ planes of the cuprate superconductors.
After extensive investigations, high temperature superconductivity with a transition temperature $T_{\rm c}$ of 56~K was 
observed in rare earth iron oxypnictides (1-1-1-1 compound)\cite{kito, ren}.
Members of superconducting iron family were observed one after another in oxygen free compounds 
such as BaFe$_2$As$_2$ having ThCr$_2$As$_2$ type crystal structure (1-2-2 compound) \cite{rotter1,rotter2}, 
LiFeAs (1-1-1 compound) \cite{lifeas} and FeSe (1-1 compound) \cite{fese}. 
New compounds having a thick perovskite layer were also reported recently \cite{shimoyama}. 
Among them, the 1-2-2 compound $AE$Fe$_2$As$_2$ ($AE$ =Ba, Ca and Sr) occupies a singular position, 
which is particularly suitable for precise physical property measurements, 
since high quality large single crystals can be grown in the congruent melting condition.
The 1-2-2 compounds have a first order antiferromagnetic AF transition on cooling at moderately high temperature, 
which always accompanied by a phase transition from tetragonal to orthorhombic structure. 
In the case of BaFe$_2$As$_2$, the crystal structure/magnetic phase transition occurs at 141~K \cite{bafe2as2}. 
The carrier doping of K for Ba, or Co for Fe strongly suppresses the transition in BaFe$_2$As$_2$.
The highest $T_{\rm c}$ of 38~K in the 1-2-2 compounds is obtained in a K doped material located adjacent to AF state \cite{rotter2}.

Another method to suppress AF ordering is applying pressure $P$. 
This has a great advantage over chemical doping since the ground state of the system can be tuned without introducing any disorder. 
Hence, many high pressure measurements have been performed in iron pnictides
~\cite{alireza,fukazawa,kotegawa1,kotegawa2,matsubayashi,ishikawa,amesp,milton1,milton2,helium,duncan}. 
However, contradicting results were reported in the 1-2-2 compound up to now. 
The simplest reason is sample quality difference. 
Polycrystalline samples are inhomogeneous and have broader phase transitions than single crystals. 
Even in single crystals, the sample growth condition makes a difference $i.e.$, the crystals grew up in Sn or self-flux. 
The contamination by other elements from the flux often influences the physical properties. 
Most reliable results are obtained in single crystals made by self-flux method. 
Even though high quality crystals were used, many groups have reported scattered and confusing results under pressure, . 
The discrepancy may be induced by the difference of the hydrostaticity in respective measurements. 

The effect of pressure inhomogeneity has been studied in CaFe$_2$As$_2$ \cite{helium} and SrFe$_2$As$_2$ \cite{kotegawa2}. 
Duncan $et~al.$ have studied the difference of hydrostaticity in BaFe$_2$As$_2$ using a piston cylinder cell, 
an alumina anvil cell with Daphne oil and a Bridgman anvil cell with solid steatite as pressure transmitting medium \cite{duncan}. 
They reported that uniaxial stress strongly suppresses structural/AF ordering and induces superconductivity at lower pressure. 

We have studied difference of hydrostaticity in very highly hydrostatic pressure range since the confusing results appear in this region
~\cite{alireza,fukazawa,matsubayashi,ishikawa,amesp,duncan}. 
We performed resistivity measurements using self-flux grown single crystals of BaFe$_2$As$_2$ from the same batch up to 16~GPa 
with a cubic anvil apparatus and compared the results with those obtained with a Bridgman anvil cell containing liquid pressure transmitting medium.

%\section{Experimental}

\begin{figure}
\includegraphics[width=8cm]{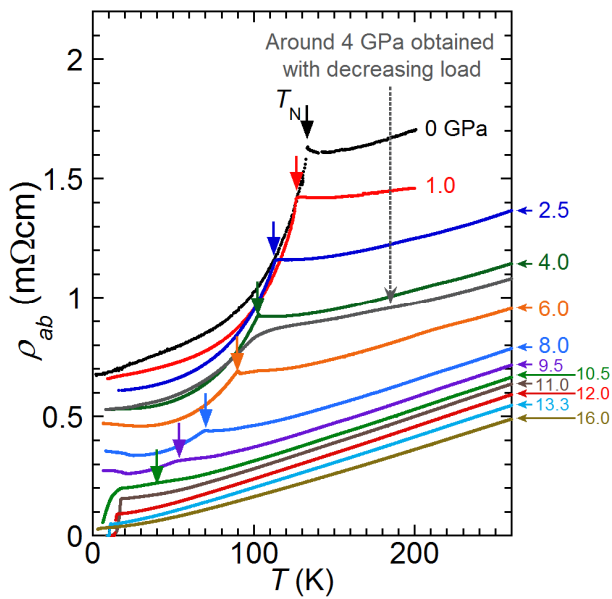}
\caption{
(color online) $T$ dependence of electrical resistivity in BaFe$_2$As$_2$ single crystal under various pressures obtained by a cubic anvil apparatus. 
%The resistivity was measured in increasing load process. 
%The arrow at $T_{\bf N}$ decreases gradually up to about 10~GPa. 
%Above 11~GPa, resistivity suddenly drops to zero, which represents appearance of superconductivity. 
In addition to the above data, resistivity around 4~GPa, which was obtained in decreasing load process, was also plotted. 
%Transition width at $T_{\bf N}$ is broadened by internal stress made by microcracks during higher pressure measurements.
}
\end{figure}

Single crystals of BaFe$_2$As$_2$ were grown by the FeAs self-flux method to avoid contamination. 
The starting materials, Ba and FeAs, were put into an alumina crucible with a ratio of 1:4 and sealed in evacuated quartz tube. 
The tube was heated up to 1140~$^\circ$C and cooled down very slowly to 1040~$^\circ$C. 
Details of the sample preparation have been previously reported \cite{Lee_sample}. 
X-ray analysis showed that the crystals had a tetragonal ThCr$_2$Si$_2$ type structure with no impurity phase. 
The cubic anvil high-pressure apparatus consists of six anvils
and homogeneously compresses the gasket cube containing a Teflon capsule~\cite{mori}. 
The sample is located inside the Teflon cell together with liquid pressure transmitting medium Daphne oil 7474~\cite{Mura1}. 
The pressure of the sample is calibrated by the resistivity anomalies of Bi, Te, Sn, and ZnS  
associated with their structural phase transitions at room temperature.
The weight loaded to the sample is kept constant during the same pressure measurement. 
This apparatus produces a very homogeneous and hydrostatic pressure up to~20 GPa. 
In the modified Bridgman cell \cite{nakanishi}, the sample is in a Teflon cell 
together with a liquid pressure- transmitting medium, a 1:1 mixture of Fluorinerts 70 and 77. 
It is expected that the solidification of Fluorinerts above 1.2 GPa yields moderate inhomogeneous pressure distributions. 
The modified Bridgman cell provides less hydrostatic condition than the cubic anvil apparatus. 
The pressure of the sample in Bridgman cell is also calibrated by the resistivity anomalies of Bi and Te at room temperature, 
and additionally from the $T_{\rm c}$ of Pb at low temperatures. 
The electrical resistivity measurements under pressure were performed by a standard dc four-probe technique with a current flow in the $ab$-plane.
The electric leads were gold wire of 20${\mu}$m in diameter with a silver-loaded epoxy resin to contact the crystal.

%\section{Results and discussion}

Figure 1 shows the temperature $T$ dependence of the electrical resistivity $\rho(T)$ of BaFe$_2$As$_2$ under pressure 
obtained using the cubic anvil apparatus. 
At ambient pressure, $\rho$($T$) decreases gradually on cooling from 300~K, 
and decreases steeply just below 134 K~as reported by many groups. 
The anomaly in the resistivity, determined by the maximum of d${\rho}$/d$T$, 
corresponds to the crystal structure and AF transition. 
We determined the transition temperature $T_{\bf N}$ from this anomaly~\cite{note}.
$T_{\bf N}$ decreases with a slope of d$T_{\bf N}$/dpressure = -7.0~K/GPa. 
The change of the transition is robust against pressure, diminishing gradually and disappearing above 10~GPa. 
Our present result below 8~GPa is similar to that obtained in a single crystal 
with a cubic anvil apparatus by Matsubayashi $et~al.$ \cite{matsubayashi}, 
and is also consistent with that previously obtained in polycrystalline samples~\cite{fukazawa}. 
An expanded view of resistivity above 10~GPa at low temperatures is shown in Fig.~2. 
$\rho(T)$ rapidly decreases below approximately 17~K and becomes zero at lower temperatures at 11~GPa, 
which indicates the collapse of structural/AF ordering and appearance of superconductivity.
A drop of the resistivity to zero appears at 11~GPa. The transition becomes sharp at 11.5~GPa.
The $T_{\rm c0}$, defined by 
temperature at which ${\rho}$ becomes zero, 
of 13~K is the highest at 11.5~GPa, deceases monotonically with increasing $P$, 
and zero resistivity was not achieved down to 3~K at 16~GPa.

\begin{figure}
\includegraphics[width=8cm]{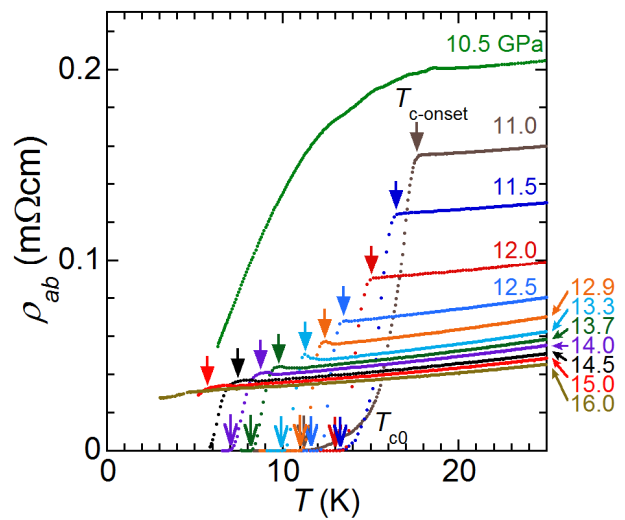}
\caption{
(color online) Expanded view of resistivity above 10~GPa at low temperature. 
%The change of resistivity around 11~GPa is very drastic as superconductivity appears. 
The onset of $T_{\rm c}$, $T_{\rm c-onset}$, is defined by anomaly in d${\rho}$/d$T$, 
and $T_{\rm c0}$ is defined by temperature at which ${\rho}$ becomes zero. 
%$T_{\rm c-onset}$ decreases monotonously with increasing pressure. 
%$T_{\rm c0}$ once increases, has a maximum around 11.5~GPa, and then decreases with increasing pressure .
} 
\end{figure}

The electrical resistivity measurement itself is insufficient, and 
further susceptibility or NMR measurements are needed to examine the bulk nature of the superconductivity.
In SrFe$_2$As$_2$, ac susceptibility measurement in a cubic anvil apparatus by Matsubayashi $et~al.$ has shown 
the appearance of bulk superconductivity in a very narrow region which is adjacent to AF phase \cite{matsubayashi}. 
The bulk nature of superconductivity in SrFe$_2$As$_2$ was also examined by $^{75}$As NMR under hydrostatic high pressure condition, 
which was performed using a soft solid Ar as pressure transmitting medium by Kitagawa $et~al.$ \cite{kitagawa}. 
We expect that the situation of BaFe$_2$As$_2$ would be similar to that of SrFe$_2$As$_2$. 

In order to study the inhomogeneity of pressure, we measured $\rho$($T$) of BaFe$_2$As$_2$ from the same batch by a modified Bridgman anvil cell. 
The crystal was aligned with its $c$-axis perpendicular to the anvil surfaces. 
As the sample was in the liquid pressure transmitting medium Fluorinerts, 
the uniaxial stress would be more or less reduced. 
However, solidification of Fluorinerts occurs above 1.2~GPa at room temperature, which reduces pressure homogeneity at higher pressure.

\begin{figure}
\includegraphics[width=8cm]{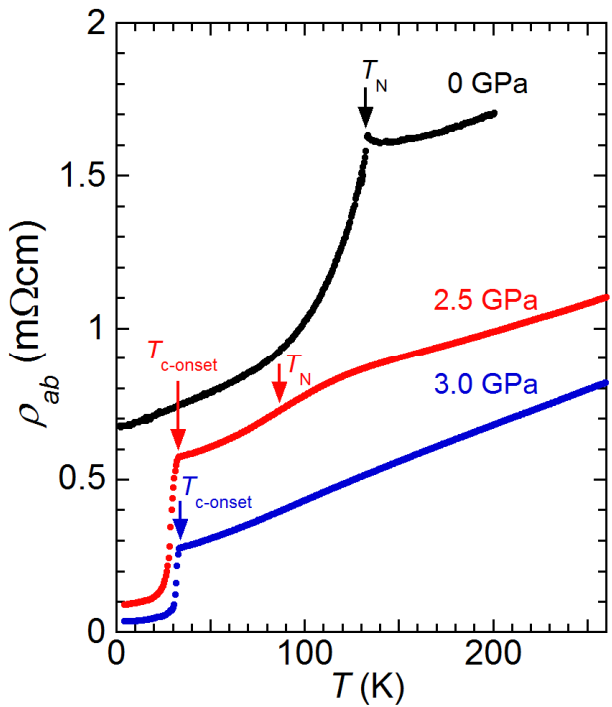}
\caption{
(color online) $T$ dependence of resistivity obtained by a modified Bridgman cell. 
%The decease of $T_{\bf N}$ is much faster than that in cubic anvil apparatus even though single crystal from same batch was used. 
}
\end{figure}

\begin{figure}
\includegraphics[width=8cm]{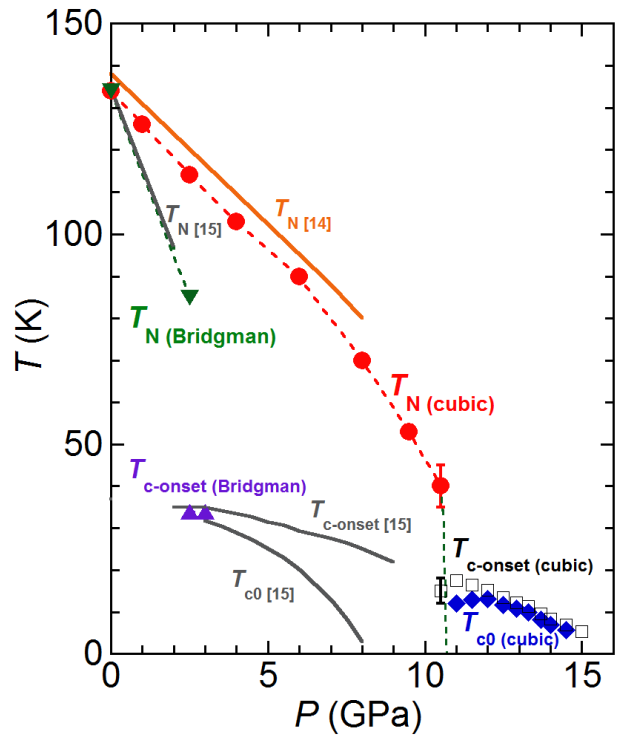}
\caption{
(color online) $P$-$T$ phase diagram of BaFe$_2$As$_2$ obtained by a cubic anvil apparatus (cubic) and a modified Bridgman anvil cell (Bridgman). 
We also plot the $T_{\rm N}$ reported by Matsubayashi $et~al.$ using a cubic anvil apparatus~\cite{matsubayashi} and
the $T_{\rm N}$, $T_{\rm c-onset}$, and $T_{\rm c0}$ reported by Ishikawa $et~al.$ using a modified Bridgman anvil cell~\cite{ishikawa}. 
%It is noted that the results obtained by cubic anvil apparatus and by Bridgman anvil cell is 
%quite different even if single crystal sample is from same batch. 
}
\end{figure}

The result obtained by the modified Bridgman cell is shown in Fig.~3. 
$T_{\bf N}$ decreases faster than that of the cubic anvil apparatus. 
Another lower $T$ anomaly appeared at 30~K above 2.5~GPa, 
which was similar to the anomaly often observed by many groups~\cite{fukazawa,ishikawa,amesp}. 
Our results at low pressure agree with the $T_{\rm c-onset}$ data obtained by Ishikawa $et~al.$ using a modified Bridgman anvil cell. 
These results are summarized and plotted in Fig.~4. 
Usually, the small inhomogeneous pressure in a modified Bridgman cell has little or no effect.
However, an inhomogeneous pressure gives rise to huge effect in the phase diagram of BaFe$_2$As$_2$, 
which is not accounted by the simple pressure inhomogeneity. 

Let us consider why BaFe$_2$As$_2$ is so sensitive to uniaxial stress. 
The X-ray measurement \cite{rotter2,su} shows that the lattice constants of BaFe$_2$As$_2$ for the $a$-, $b$-, and $c$-axes 
($a$, $b$, $c$) have a characteristic feature. 
The $T$ dependence of $a$, $b$, and $c$ are highly anisotropic and completely different in the tetragonal and the orthorhombic structures. 
In the tetragonal structure, $c$ decreases and $a$ ($b$) remains constant with decreasing $T$. 
On the contrary, in the orthorhombic structure the average of $a$ and $b$ decreases, and $c$ becomes constant with decreasing $T$ $i.e.$, 
the volume reduction in the tetragonal structure occurs through the shrinkage along $c$-axis, 
and that in orthorhombic structure through shrinkage along the $a$- and $b$-axes on cooling. 
A similar result for the $T$ dependence of lattice parameters is observed in thermal expansion \cite{sergey}. 
Then we consider the situation where uniaxial stress is applied to the crystal along the $c$-axis, which is plotted schematically in Fig.~5. 
When uniaxial stress is applied, the free energy is minimized by the crystal shrinking along the $c$-axis. 
The tetragonal structure is stabilized by uniaxial stress, which explains the high sensitivity to the inhomogeneity of pressure. 
In this way, a small amount of ``tetragonal structure phase induced by inhomogeneous pressure" develops 
as small island in the sea of orthorhombic structure/AF phase in the lower pressure region. 
Superconductivity appears in the induced tetragonal structure phase at $T_{\rm c}$ probably determined from so-called ``Lee-plot" 
which is  empirical relation between $T_{\rm c}$ and As-Fe-As bond angle \cite{lee}. 
As seen in Fig.~4, the $T_{\rm c-onset}$ is smooth having a maximum value at around 38~K at low pressure 
and finally seems to disappear around 16~GPa, 
which is similar to the $P$ dependence of $T_{\rm c}$ observed in Ba$_{0.6}$K$_{0.4}$Fe$_2$As$_2$ \cite{takeshita}. 
Indeed, the tendency of the As-Fe-As bond angle moving away from the highest-$T_{\rm c}$ angle with increasing $P$ is 
already reported by Kimber $et~al$.~\cite{kimber}

%\section{Conclusion}

In conclusion, we have measured the electrical resistivity of single-crystal BaFe$_2$As$_2$ 
under high pressure with a cubic anvil apparatus and with a modified Bridgman anvil cell to determines the pressure-$T$ phase diagram.
Under hydrostatic conditions, the crystal structure/AF transition temperature decreases robustly against pressure 
and disappears around 10~GPa. 
Superconductivity appears in the pressure range between 11 and 14~GPa. 
The highest $T_{\rm c0}$ of 13~K is  obtained at 11.5~GPa, which is probably adjacent to the orthorhombic/AF phase, 
and $T_{\rm c}$ decreases monotonically with increasing $P$. 
Uniaxial stress, which is induced by tiny departure from hydrostaticity, strongly suppresses structural/AF ordering 
and stabilizes superconductivity.
In order to establish the $P$-$T$ phase diagram of BaFe$_2$As$_2$, 
more precise structural study under pressure in which pressure condition is the same as that 
utilized in resistivity or magnetic susceptibility measurements is desirable. 

\begin{figure}
\includegraphics[width=8cm]{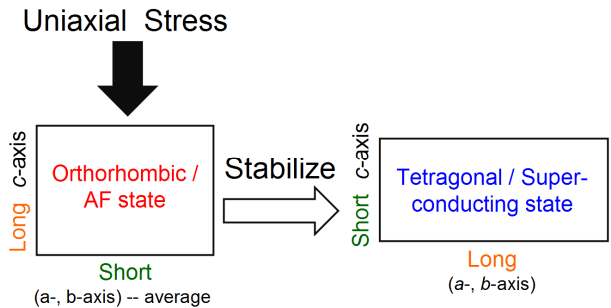}
\caption{
(color online) Schematic view of role of uniaxial stress in BaFe$_{2}$As$_{2}$.
%Tetragnal structure is stabilized by uniaxial stress.  
}
\end{figure}

The authors thank C. Terakura, T. Saito, F. Ishikawa and Y. Yamada for valuable discussions and suggestions. 
They also thank D. E. MacLaughlin for his thorough reading of this manuscript. 
This work is supported by Grants-in-Aid for Scientific Research from the Ministry of Education, Culture, Sports, Science and Technology (MEXT) 
and the Japan Society for the Promotion of Science (JSPS), and Innovative Areas ``Heavy Electrons" (No. 20102005) from MEXT, 
Global COE program of Chiba University.

\newpage

\begin{thebibliography}{99}


\bibitem[*]{email}Corresponding author, \\
E-mail: kohori@faculty.chiba-u.jp

\bibitem{kamihara} Y. Kamihara, T. Watanabe, M. Hirano, and H. Hosono, J. Am. Chem. Soc. {\bf 130}, 3296 (2008). 

\bibitem{kito} H. Kito, H. Eisaki, and Y. Iyo, J. Phys. Soc. Jpn. {\bf 77}, 063707 (2008). 

\bibitem{ren} Z.-A. Ren, J. Yang, W. Lu, W. Yi, X.-L. Shen, Z.-C.-Li, G. C. Che, X.-L. Dong, L.-L. Sun, F. Z-X. Zhao
Europhs. Lett. {\bf 82}, 57002 (2008). 

\bibitem{rotter1} M. Rotter, M. Tegel, and D. Johrendt, Phys. Rev. Lett. {\bf 101}, 107006 (2008). 

\bibitem{rotter2} M. Rotter, M. Tegel, and D. Johrendt, I. Schellenberg, W. Hermes and R. P\"{o}ttgen, Phys. Rev. {\bf B 78}, 020503(R) (2008).

\bibitem{lifeas} X. C. Wang, Q. Q. Liu, Y. X. Lv, W. B. Gao, L. X. Yang, R. C. Yu, F. Y. Li and C. Q. Jin, Solid State Commun. {\bf 148}, 538 (2008).

\bibitem{fese} F.-C. Hsu, J.-Y. Luo, K.-W. Yeh, T.-K. Chen, T.-W. Huang, P.-M. Wu, Y.-C. Lee, Y.-L. Huang, Y.-Y. Chu, D.-C. Yan, 
and M.-K. Wu, Proc. Natl. Acad. Sci. USA {\bf 105}, 14262 (2008).

\bibitem{shimoyama} H. Ogino, Y. Matsumura, Y. Katsura, K. Ushiyama, S. Horii, K. Kishio, and J. Shimoyama, 
Suppercond. Sci. Technol. {\bf 22}, 075008 (2009). 

\bibitem{bafe2as2} Q. Huang, Y. Qiu, W. Bao, M.A. Green, J. W. Lynn, Y. C. Gasparovic, T. Wu, G. Wu, and X. H. Chen, 
Phys. Rev. Lett. {\bf 101}, 257003 (2008). 

\bibitem{alireza} P. L. Alireza, Y. T. C. Ko, J. Gillett, C. M. Petrone, J. M. Cole, G. G. Lonzarich and S. E Sebastian, 
J. Phys.: Condens. Matter, {\bf 21}, 012208 (2008). 

\bibitem{fukazawa} H. Fukazawa, N. Takeshita, T. Yamazaki, K. Kondo, K. Hirayama, Y. Kohori, K. Miyazawa, H. Kito, H. Eisaki, and A. Iyo, 
J. Phys. Soc. Jpn. {\bf 77}, 105004 (2008).

\bibitem{kotegawa1} H. Kotegawa, H. Sugawara, and H. Tou, J. Phys. Soc. Jpn. {\bf 78}, 013709 (2009).

\bibitem{kotegawa2} H. Kotegawa, T. Kawazoe, H.Sugawara, K. Murata, and H. Tou, J. Phys. Soc. Jpn. {\bf 78}, 083702 (2009).

\bibitem{matsubayashi} K. Matsubayashi, N. Katayama, K. Ohgushi, A. Yamada, K. Munakata, T. Matsumoto, and Y. Uwatoko, 
J. Phys. Soc. Jpn. {\bf 78}, 073706 (2009).

\bibitem{ishikawa} F. Ishikawa, N. Eguchi, M. Kodama, K. Fujimaki, M. Einaga, A. Ohmura, A. Nakayama, A. Mitsuda, Y. Yamada, 
Phys. Rev. {\bf B 79}, 172506 (2009).

\bibitem{amesp} E. Colombier, S. L. Budfko, N. Ni, and P. C. Canfield, Phys. Rev. {\bf B 79}, 224518 (2009).

\bibitem{milton1}  M. S. Torikachvili, S. L. Budfko, N. Ni, and P. C. Canfield, Phys. Rev. Lett. {\bf 101}, 057006 (2008).

\bibitem{milton2} M. S. Torikachvili, S. L. Budfko, N. Ni, and P. C. Canfield, Phys. Rev. {\bf B 78}, 104527 (2008).

\bibitem{helium} W. Yu, A. A. Aczel, T. J. Williams, S. L. Budfko, N. Ni, P. C. Canfield, and G. M. Luke , Phys. Rev. {\ B 79}, 020511(R) (2009).

\bibitem{duncan} W. J. Duncan, O. P. Welzel, C. Harrison, X. F. Wang, X. H. Chen, F. M. Grosche and P. G. Niklowitz, 
J. Phys.: Condens. Matter {\bf 22}, 052201 (2010).

\bibitem{Lee_sample} C. H. Lee, K. Kihou, K. Horigane, S. Tsutsui, T. Fukuda, H. Eisaki, A. Iyo, H. Yamaguchi, A. Q. R. Baron, M. Braden, 
and K. Yamada, J. Phys. Soc. Jpn. {\bf 79}, 014714 (2010).

\bibitem{mori}  N. M\^{o}ri, H. Takahashi, and N. Takeshita: High Pressure Res. {\bf 24}, 225 (2004). 

\bibitem{Mura1} K. Murata, K. Yokogawa, H. Yoshino, S. Klotz, P. Munsch, A. Irizawa, M. Nishiyama, K. Iizuka, T. Nanba, T. Okada, Y. Shiraga, 
and S. Aoyama, Rev. Sci. Instrum. {\bf 79}, 085101 (2008).

\bibitem{note} We express the structural/AF transition as $T_{\rm N}$ since the structural and AF transitions 
coincides with each other in almost all the 1-2-2 compounds. 
Therefore, we also change the notation of $T_{\rm S}$~\cite{matsubayashi} and $T_{\rm SDW}$~\cite{ishikawa} as  $T_{\rm N}$ in this paper. 
Note that the definition of these anomalies itself from experimental $\rho$ is the same. 

\bibitem{nakanishi}  T. Nakanishi, N. Takeshita, and N. M\^{o}ri, Rev. Sci. Instrum. {\bf 73}, 1828 (2002). 

\bibitem{kitagawa} K. Kitagawa, N. Katayama, H. Gotou, T. Yagi, K. Ohgushi, T. Matsumoto, Y. Uwatoko, and M. Takigawa, 
Phys. Rev. Lett. {\bf 103}, 257002 (2009).

\bibitem{su} Y. Su, P. Link, A. Schneidewind, Th. Wolf, P. Adelmann, Y. Xiao, M. Meven, R. Mittal, M. Rotter, D. Johrendt Th. Bruecke and M. Loewenhaupt,
Phys. Rev. {bf 79}, 064504 (2009).

\bibitem{sergey} Sergey L. Bud'ko, Ni Ni, Paul C. Canfield, arXiv:0907.2936.

\bibitem{lee} C. H. Lee, A. Iyo, H. Eisaki, H. Kito, M. T. Fernandez-Diaz, T. Ito, K. Kihou, H. Matsuhata, M. Braden, and K. Yamada, 
J. Phys. Soc. Jpn. {\bf 77}, 083704 (2008). 

\bibitem{kimber} S. A. J. Kimber, A. Kreyssig, Y.-Z. Zhang, H. O. Jeschke, R. Valenti, F. Yokaichiya, E. Colombier, J. Yan, T. C. Hansen, T. Chatterji, 
R. J. McQueeney, P. C. Canfield, A. I. Goldman, and D. N. Argyriou, Nature Matter. {\bf 8}, 471 (2009). 

\bibitem{takeshita} T. Yamazaki, N. Takeshita, K. Kondo, R. Kobayashi, Y. Yamada, H. Fukazawa, Y. Kohori, P. M. Shirage, K. Kihou, H. Kito, 
H. Eisaki, and A. Iyo, Proceeding of AIRAPT-22~\verb|&|~HPCJ-50, to be published in J. Phys: Conf. Ser. (2010). 




\end{thebibliography}
\end{document}